\documentclass[12pt]{article}
\usepackage{amsfonts,amssymb,amsmath,mathrsfs}
\usepackage{hyperref}
\newcommand{\beq}{\begin{equation}}
\newcommand{\eeq}{\end{equation}}
\newcommand{\vp}{\vphantom}
\newcommand{\hp}{\hphantom}
\begin{document}
\begin{center}
{\Large\bf On integrability of massless $AdS_4\times\mathbb{CP}^3$ superparticle equations}\\[0.5cm]
{\large D.V.~Uvarov\footnote{E-mail: d\_uvarov@\,hotmail.com}}\\[0.2cm]
{\it NSC Kharkov Institute of Physics and Technology,}\\ {\it 61108 Kharkov, Ukraine}\\[0.5cm]
\end{center}
\begin{abstract}
Lax representation is elaborated for the equations of motion of massless superparticle on the $AdS_4\times\mathbb{CP}^3$ superbackground that proves their classical integrability.
\end{abstract}

\section{Introduction}

Discovery of the integrability of $AdS_5\times S^5$ superstring equations \cite{BPR} and of the dilatation operator in the dual $D=4$ $\mathcal N=4$ super-Yang-Mills theory \cite{Minahan} determined the direction of major progress in studying the $AdS_5/CFT_4$ correspondence \cite{Maldacena} over the last decade.\footnote{For a recent collection of reviews see \cite{Beisert}.} The situation with the Aharony-Bergman-Jafferis-Maldacena (ABJM) correspondence \cite{ABJM} is more intricate. Conjectured gravity dual of $D=3$ $\mathcal N=6$ superconformal $U(N)_k\times U(N)_{-k}$ gauge theory reduces to a string theory only in the special sublimit of the 't Hooft limit defined by the conditions $k^5\gg N\gg1$. Moreover this IIA superstring theory 'lives' on $AdS_4\times\mathbb{CP}^3$ superbackground \cite{Watamura} preserving 24 of 32 space-time supersymmetries contrary to the IIB $AdS_5\times S^5$ superbackground that is maximally supersymmetric. Non-maximal supersymmetry on both sides of ABJM duality significantly complicates its exploration (see \cite{Klose} for a review).

Thus application of the supercoset approach, suggested in \cite{MT98} to construct the $AdS_5\times S^5$ superstring action as a $2d$ sigma-model on $PSU(2,2|4)/(SO(1,4)\times SO(5))$ supercoset manifold, gives only part of the $AdS_4\times\mathbb{CP}^3$ superstring action since only a $(10|24)-$dimensional subspace of the $AdS_4\times\mathbb{CP}^3$ superspace is isomorphic to the $OSp(4|6)/(SO(1,3)\times U(3))$ supercoset manifold \cite{AF}, \cite{Stefanski}.\footnote{Alternative approach to constructing the $OSp(4|6)/(SO(1,3)\times U(3))$ sigma-model action relies on the introduction of the pure spinor variables \cite{PS}.} These are 24 Grassmann coordinates of the $OSp(4|6)/(SO(1,3)\times U(3))$ supercoset manifold that are in one-to-one correspondence with the supersymmetries of the $AdS_4\times\mathbb{CP}^3$ superbackground so that construction of the complete $AdS_4\times\mathbb{CP}^3$ superstring action \cite{GSWnew} requires extension of the $OSp(4|6)/(SO(1,3)\times U(3))$ sigma-model by 8 fermionic fields associated with the broken supersymmetries.\footnote{The $OSp(4|6)/(SO(1,3)\times U(3))$ sigma-model arises upon partial $\kappa-$symmetry gauge fixing of the $AdS_4\times\mathbb{CP}^3$ superstring by setting to zero 8 Grassmann coordinates related to broken supersymmetries. This constrains its range of application \cite{AF}, \cite{GSWnew}.} To this end more roundabout way is to be taken that relies on double-dimensional reduction \cite{DHIS}, \cite{Howe} of the $D=11$ supermembrane on $AdS_4\times S^7$ superbackground \cite{deWit}. Since this background is maximally supersymmetric the supermembrane action is constructed using the generalization of the supercoset approach \cite{MT98}. Another necessary ingredient is the Hopf fibration realization of the 7-sphere $S^7=\mathbb{CP}^3\times S^1$ \cite{Nilsson}, \cite{STV} whose $S^1$ fiber is identified with the world-volume compact dimension in the process of double-dimensional reduction. Resultant $AdS_4\times\mathbb{CP}^3$ superstring action has rather complicated highly non-linear structure that does not simplify enough even after fixing the $\kappa-$symmetry gauge \cite{GrSW}, \cite{U09} to address directly issues of quantization and spectrum identification.

At the same time $PSU(2,2|4)/(SO(1,4)\times SO(5))$ and $OSp(4|6)/(SO(1,3)\times U(3))$ sigma-models are known to belong to a class of classically integrable $2d$ sigma-models on supercoset manifolds with $\mathbb{Z}_4-$graded isometry superalgerbas \cite{Adam}, \cite{BSZ}, \cite{STWZ} so one may hope to establish integrability of the $AdS_4\times\mathbb{CP}^3$ superstring equations following from the complete action by properly extending the Lax representation of the $OSp(4|6)/(SO(1,3)\times U(3))$ sigma-model equations \cite{AF}, \cite{Stefanski} by contributions of 8 'broken' fermions. The first steps in this direction were made in \cite{SW10}, \cite{CSW} where
the $OSp(4|6)/(SO(1,3)\times U(3))$ sigma-model Lax connection was extended by linear and quadratic terms in the 'broken' fermions in such a way that its curvature turns to zero up to quadratic order in them on the superstring equations truncated to the same order. Even at this order the expression for the extended Lax connection appears rather involved. That is why in \cite{U-IJMPA} we suggested to use the $\kappa-$symmetry gauge freedom to retain in the sector of broken supersymmetries a part of coordinates the complete expansion in which of the Lax connection can be recovered with less efforts and then by gradually relaxing the gauge condition the dependence on other coordinates may be examined.

To simplify the problem as much as possible one can initially
concentrate on the zero-mode sector described by the massless
superparticle model on the $AdS_4\times\mathbb{CP}^3$
superbackground. In Ref.~\cite{U-NPB} after working out the Lax
representation for equations of motion of the massless
$OSp(4|6)/(SO(1,3)\times U(3))$ superparticle \cite{Stefanski} we
extended it to include contributions of 4 Grassmann coordinates
related to the broken part of $D=3$ $\mathcal N=8$ Poincare
supersymmetry while other 4 coordinates related to the broken part
of conformal supersymmetry were gauged away. Here we relax this
partial $\kappa-$symmetry gauge condition and prove integrability
of the full-fledged $AdS_4\times\mathbb{CP}^3$ superparticle
equations. Namely we show that they admit the Lax representation
\beq\label{lax1} \frac{d\mathscr L}{d\tau}+[\mathscr M,\mathscr
L]=0, \eeq where $\mathscr M$ is the world-line projection of the
left-invariant $osp(4|6)$ Cartan forms definition and $\mathscr L$
can be presented as the $osp(4|6)$ superalgebra valued
differential operator acting on the superparticle action
functional \beq\label{lax2compl}
\begin{array}{rl}
\mathscr L=&\left(M_{0'm}\frac{\partial}{\partial G_{\tau}\vp{G}^{0'}\vp{G}_m}+\frac12D\frac{\partial}{\partial\Delta_{\tau}}-M_{mn}\frac{\partial}{\partial G_{\tau mn}}-M_{3m}\frac{\partial}{\partial G_{\tau 3m}}+T_a\frac{\partial}{\partial\Omega_{\tau a}}+T^a\frac{\partial}{\partial\Omega_{\tau}\vp{\Omega}^a}\right.\\[0.2cm]
+&\left.\widetilde V_a{}^a\frac{\partial}{\partial\vp{\hat{\widetilde\Omega}}\widetilde\Omega_{\tau b}{}^b}-\frac14Q^{\vp{a}}_{(1)}\vp{Q}^a_\mu\frac{\partial}{\partial\bar\omega^{\vp{a}}_{(3)}\vp{\bar\omega}^{\hp{\tau}a}_{\tau\mu}}+\frac14\bar Q_{(1)\mu a}\frac{\partial}{\partial\omega_{(3)\tau\mu a}}+\frac14Q^{\vp{a}}_{(3)}\vp{Q}^a_\mu\frac{\partial}{\partial\bar\omega_{(1)}\vp{\bar\omega}^{\hp{\tau}a}_{\tau\mu}}-\frac14\bar Q_{(3)\mu a}\frac{\partial}{\partial\omega_{(1)\tau\mu a}}\right)S
\end{array}
\eeq
as was proposed in \cite{U-NPB}.

Organization of the paper is the following. In Section 2 necessary data on the geometry of $AdS_4\times\mathbb{CP}^3$ superspace is briefly reviewed and then in Section 3 equations of motion for the $AdS_4\times\mathbb{CP}^3$ superparticle are derived and their Lax representation is worked out.

\section{$osp(4|6)$, $osp(4|8)$ Cartan forms and $AdS_4\times\mathbb{CP}^3$ supervielbein}

Expressions for the geometric constituents of the
$AdS_4\times\mathbb{CP}^3$ superspace were obtained in
Ref.~\cite{GSWnew}. In this Section to make the presentation
self-contained we review the construction of the supervielbein
bosonic components putting emphasis on the realization of
$osp(4|6)$ and $osp(4|8)$ isometry superalgerbas of
$AdS_4\times\mathbb{CP}^3$ and $AdS_4\times S^7$ superbackgrounds
as $D=3$ $\mathcal N=6,8$ superconformal algebras of dual ABJM
gauge theory \cite{ABJM}.

Left-invariant $osp(4|6)$ Cartan forms in the conformal basis admit the following decomposition \cite{U08}
\beq
\label{osp46cf}
\begin{array}{rl}
\mathscr C(d)=\mathscr G^{-1}d\mathscr G=&\Delta(d)D+
\omega^m(d)P_m+c^m(d)K_m
+G^{mn}(d)M_{mn}\\[0.2cm]
+&\Omega_a(d)T^a+\Omega^a(d)T_a+\widetilde\Omega_a{}^b(d)\widetilde V_b{}^a+\widetilde\Omega_b{}^b(d)\widetilde V_a{}^a\\[0.2cm]
+&\omega^\mu_a(d)Q^a_\mu+\bar\omega^{\mu a}(d)\bar Q_{\mu a}+\chi_{\mu
a}(d)S^{\mu a}+\bar\chi^a_\mu(d)\bar S^\mu_a
\end{array}
\eeq
over the generators of $D=3$ conformal algebra $(D, P_m, K_m, M_{mn})$, $su(4)\sim so(6)$ generators divided into the generators $\widetilde V_b{}^a$ of the $U(3)$ stability group of $\mathbb{CP}^3=SU(4)/U(3)$ manifold and the coset generators $(T_a, T^a)$, and fermionic generators of $D=3$ $\mathcal N=6$ Poincare $(Q^a_\mu, \bar Q_{\mu a})$ and conformal $(S^{\mu a}, \bar S^\mu_a)$ supersymmetries that carry $SL(2,\mathbb{R})$ spinor index $\mu=1,2$ and $SU(3)$ (anti)fundamental representation index $a=1,2,3$ in accordance with the decomposition $\mathbf 6=\mathbf 3\oplus\bar{\mathbf 3}$ of the $SO(6)$ vector representation on $SU(3)$ representations.

To characterize the geometry of $OSp(4|6)/(SO(1,3)\times U(3))$ supermanifold Cartan forms related to the generators $g_{(\mathrm k)}$ with definite eigenvalues $i^{\mathrm k}$ under the $\mathbb Z_4$ automorphism of the $osp(4|6)$ superalgebra are used. Corresponding form of (\ref{osp46cf}) is
\beq\label{gradedcf}
\mathscr C(d)=\mathscr C_{(0)}(d)+\mathscr C_{(2)}(d)+\mathscr C_{(1)}(d)+\mathscr C_{(3)}(d),
\eeq
where
\beq
\begin{array}{rl}
\mathscr C_{(0)}(d)=&2G^{3m}(d)M_{3m}+G^{mn}(d)M_{mn}+\widetilde\Omega_a{}^b(d)\widetilde V_b{}^a+\widetilde\Omega_a{}^a(d)\widetilde V_b{}^b\in g_{(0)},\\[0.2cm]
\mathscr C_{(2)}(d)=&2G^{0'm}(d)M_{0'm}+\Delta(d)D+\Omega_a(d)T^a+\Omega^a(d)T_a\in g_{(2)},\\[0.2cm]
\mathscr C_{(1)}(d)=&\omega_{(1)}\vp{\omega}^\mu_{a}(d)Q_{(1)}\vp{Q}^a_\mu+\bar\omega^{\vp{\mu c}}_{(1)}\vp{\bar\omega}^{\mu a}(d)\bar Q_{(1)\mu a}\in g_{(1)},\\[0.2cm]
\mathscr C_{(3)}(d)=&\omega_{(3)}\vp{\omega}^\mu_{a}(d)Q_{(3)}\vp{Q}^a_\mu+\bar\omega^{\vp{\mu c}}_{(3)}\vp{\bar\omega}^{\mu a}(d)\bar Q_{(3)\mu a}\in g_{(3)}.
\end{array}
\eeq
Definite eigenvalues under the $\mathbb Z_4$ isomorphism have the $so(2,3)$ generators
\beq
M_{0'm}=\frac12(P_m+K_m),\quad M_{0'3}=-D,\quad M_{3m}=\frac12(K_m-P_m)
\eeq
and fermionic generators
\beq\label{gradedfermidef}
Q^{\vp{a}}_{(1)}\vp{Q}^a_\mu=Q^a_\mu+iS^a_\mu,\quad\bar Q_{(1)}\vp{Q}_{\mu a}=\bar Q_{\mu a}-i\bar S_{\mu a};\quad Q^{\vp{a}}_{(3)}\vp{Q}^a_\mu=Q^a_\mu-iS^a_\mu,\quad\bar Q_{(3)}\vp{Q}_{\mu a}=\bar Q_{\mu a}+i\bar S_{\mu a}.
\eeq
Bosonic and fermionic Cartan forms from $\mathscr C_{(1,2,3)}$ eigenspaces
\beq\label{osp46bviel}
G^{0'm}(d)=\frac12(\omega^m(d)+c^m(d)),\quad\Delta(d),\quad\Omega_a(d),\quad\Omega^a(d)
\eeq
and
\beq
\omega_{(1)}\vp{\omega}^\mu_{a}(d)=\frac12(\omega^\mu_a(d)+i\chi^\mu_a(d)),\quad\omega_{(3)}\vp{\omega}^\mu_{a}(d)=\frac12(\omega^\mu_a(d)-i\chi^\mu_a(d))
\eeq
and c.c. are identified with the $OSp(4|6)/(SO(1,3)\times U(3))$ supervielbein bosonic and fermionic components, other Cartan forms
\beq
G^{3m}(d)=-\frac12(\omega^m(d)-c^m(d)),\quad G^{mn}(d),\quad\widetilde\Omega_a{}^b(d)
\eeq
describe the $SO(1,3)\times U(3)$ connection.

Geometric constituents of the $OSp(4|8)/(SO(1,3)\times SO(7))$ supermanifold are constructed out of the $osp(4|8)$ Cartan forms that in conformal basis read \cite{U09}
\beq\label{osp48cf}
\begin{array}{rl}
\hat{\mathscr G}^{-1}d\hat{\mathscr G}=&\underline\Delta(d)D+\underline{\omega}^m(d)P_m+\underline c^m(d)K_m+\underline G^{mn}(d)M_{mn}\\[0.2cm]
+&\Omega_a(d)T^a+\Omega^a(d)T_a+\widetilde\Omega_a(d)\widetilde T^a+\widetilde\Omega^a(d)\widetilde T_a\\[0.2cm]
+&\widetilde\Omega_a{}^b(d)\widetilde V_b{}^a+\widetilde\Omega_b{}^b(d)\widetilde
V_a{}^a+\Omega_a{}^4(d)V_4{}^a+\Omega_4{}^a(d)V_a{}^4+h(d)H\\[0.2cm]
+&\underline\omega^\mu_a(d)Q^a_\mu+\underline{\bar\omega}{}^{\mu a}(d)\bar Q_{\mu a}+\underline\chi{}_{\mu a}(d)S^{\mu a}+\underline{\bar\chi}{}^a_\mu(d)\bar S^\mu_a\\[0.2cm]
+&\omega^\mu_4(d)Q^4_\mu+\bar\omega^{\mu 4}(d)\bar Q_{\mu 4}+\chi_{\mu 4}(d)S^{\mu 4}+\bar\chi^4_\mu(d)\bar S^\mu_4.
\end{array}
\eeq
There can be chosen $OSp(4|8)/(SO(1,3)\times SO(7))$ representative suitable for the reduction to 10 dimensions
\beq\label{scoset}
\hat{\mathscr G}=\mathscr Ge^{yH}\mathscr G_{\mathrm{br}},
\eeq
where coordinate $y\in[0,2\pi)$ parametrizes $S^1$ fiber and $\mathscr G_{\mathrm{br}}$ is a function of 8 Grassmann coordinates for the broken supersymmetries $\upsilon_\mu=(\theta_\mu, \bar\theta_\mu, \eta_\mu, \bar\eta_\mu)$. There have been underlined those of the $osp(4|6)$ Cartan forms that acquire dependence on $dy$, $\upsilon$ and $d\upsilon$ in addition to that on coordinates of the $OSp(4|6)/(SO(1,3)\times U(3))$ supercoset manifold.\footnote{Note that such a choice of the $OSp(4|8)/(SO(1,3)\times SO(7))$ element rules out dependence of the $osp(4|8)$ Cartan forms on $y$ itself.} $so(8)$ generators in (\ref{osp48cf}) have been transformed to the basis adapted to the Hopf fibration realization of the 7-sphere. In contrast to the consideration of Ref.~\cite{GSWnew} we manifestly decomposed $so(8)$ generators on the $SO(6)$ irreducible components and transformed them into the $\mathbf 3\oplus\bar{\mathbf 3}$ basis. In addition we proposed convenient realization for the K\" ahler 2-form on $\mathbb{CP}^3$ manifold so that the generators $(\widetilde V_a{}^b, T_a, T^a)$ span its $su(4)$ isometry algebra and commute with the $S^1$ generator $H$ \cite{U09}, \cite{U10}, \cite{U-IJMPA}. Remaining 12 generators $(\widetilde T_a, \widetilde T^a, V_a{}^4, V_4{}^a)$ belong to the $so(8)/(su(4)\times u(1))$ coset. This choice of the K\" ahler tensor also diagonalizes two projectors \cite{Nilsson}, \cite{GSWnew} that divide 32 fermionic generators of $osp(4|8)$ superalgebra (and associated coordinates) into 24 generators of $osp(4|6)$ superalgebra and 8 generators corresponding to the supersymmetries broken by the $AdS_4\times\mathbb{CP}^3$ superbackground. In conformal basis these generators are $(Q^a_\mu, \bar Q_{\mu a}; S^{\mu a}, \bar S^\mu_a)$ and $(Q^4_\mu, \bar Q_{\mu 4}; S^{\mu 4}, \bar S^\mu_4)$ respectively in accordance with the decomposition $\mathbf 4=\mathbf 3\oplus\mathbf 1$, $\bar{\mathbf 4}=\bar{\mathbf 3}\oplus\bar{\mathbf 1}$ of the (anti)fundamental representation of $SU(4)$ on $SU(3)$ representations.

As a result the $AdS_4\times S^7$ supervielbein bosonic components have the following expression in terms of $osp(4|8)$ Cartan forms 
\beq\label{ads4s7viel}
\begin{array}{c}
\hat E^{m'}(d)=\left(\frac12(\underline{\omega}^m(d)+\underline c^m(d)),\ -\underline\Delta(d)\right),\quad\hat E^{11}(d)=h(d)+\widetilde\Omega_a{}^a(d)\\[0.2cm]
E_{a}(d)=i(\Omega_a(d)+\widetilde\Omega_a(d)),\quad E^{a}(d)=i(\Omega^a(d)+\widetilde\Omega^a(d))
\end{array}
\eeq
generalizing (\ref{osp46bviel}). To perform the reduction to $10$ dimensions it is necessary to single out $dy-$dependent contributions in (\ref{ads4s7viel}). Because of the form of (anti)commutation relations of $osp(4|8)$ superalgebra such terms appear only in the supervielbein components tangent to Anti-de Sitter part of the background and $S^1$
\beq\label{ads4s7viel'}
\hat E^{m'}(d)=G^{m'}(d)+G^{m'}_ydy,\quad\hat E^{11}(d)=\Phi dy+a(d).
\eeq
Since $G_y^{m'}\not=0$ the form of $\hat E^{m'}(d)$ deviates from the Kaluza-Klein ansatz \cite{DHIS}, \cite{Howe} so that the $SO(1,4)$ tangent space Lorentz rotation should be applied to remove the contributions proportional to $dy$
\beq\label{adscpviel}
\begin{array}{rl}
(\mathsf L\hat E)^{m'}(d)=&\mathsf L^{m'}{}_{n'}\hat E^{n'}+\mathsf L^{m'}{}_{11}\hat E^{11}=E^{m'}(d),\\[0.2cm]
(\mathsf L\hat E)^{11}(d)=&\mathsf L^{11}{}_{m'}\hat E^{m'}+\mathsf L^{11}{}_{11}\hat E^{11}=\Phi_L(dy+A_L(d)).
\end{array}
\eeq
$E^{m'}(d)$ is identified with the
$AdS_4\times\mathbb{CP}^3$ supervielbein components tangent to
$AdS_4$ space-time,
\beq
\Phi_L=\sqrt{\Phi^2+G_y^2},\quad
G^2_y=G_{ym'}G_y^{m'}=G_y\cdot G_y
\eeq
determines the $D=10$ dilaton
superfield $\phi(\upsilon)=3/2\log\Phi_L$ and $A_L$ -- RR 1-form potential.
The components of the Lorentz rotation matrix
\beq\label{Lrot}
||\mathsf L||=\left(
\begin{array}{rl}
\mathsf L^{m'}{}_{n'} & \mathsf L^{m'}{}_{11}\\[0.2cm]
\mathsf L^{11}{}_{m'} & \mathsf L^{11}{}_{11}
\end{array}
\right)\in SO(1,4),
\eeq
defined by the condition $\mathsf L^{m'}{}_{n'}G^{n'}_y+\mathsf L^{m'}_{11}\Phi=0$ read
\beq\label{Lrot2}
\begin{array}{rl}
\mathsf L^{m'}{}_{n'}=&\delta^{m'}_{n'}+\frac{\Phi-\Phi_L}{\Phi_LG^{2\vp{\hat b}}_y}G^{m'}_yG_{yn'},\quad
\mathsf L^{m'}{}_{11}=-\Phi_L^{-1}G^{m'}_y,\\[0.2cm]
\mathsf L^{11}{}_{m'}=&\Phi_L^{-1}G_{ym'},\quad \mathsf L^{11}{}_{11}=\Phi_L^{-1}\Phi.
\end{array}
\eeq
Other $AdS_4\times S^7$ supervielbein bosonic components $E_{a}(d)$ and $E^{a}(d)$ do not depend on $dy$ and can be directly identified with the $AdS_4\times\mathbb{CP}^3$ supervielbein components tangent to the $\mathbb{CP}^3$ manifold.

\section{Massless $AdS_4\times\mathbb{CP}^3$ superparticle}

Massless $AdS_4\times\mathbb{CP}^3$ superparticle action arises in the tension-to-infinity limit of the $AdS_4\times\mathbb{CP}^3$ superstring \cite{GSWnew} or in the mass-to-zero limit of the $D0-$brane \cite{GrSW} and is constructed out of the world-line pullbacks of supervielbein bosonic components discussed in the previous Section
\beq\label{action}
S=\int\frac{d\tau}{e}\Phi_L\left(E_{\tau m'}E^{m'}_\tau-E_{\tau a}E_\tau\vp{E}^a\right).
\eeq
Note that when the coordinates related to the generators of broken supersymmetries are set to zero by using the $\kappa-$symmetry above action functional reduces to that of the massless superparticle on the $OSp(4|6)/(SO(1,3)\times U(3))$ supermanifold \cite{Stefanski}. Substituting expressions (\ref{ads4s7viel'}), (\ref{adscpviel}), (\ref{Lrot2}) that define the form of the supervielbein components tangent to $AdS_4$ allows to rewrite (\ref{action}) as
\beq\label{action'}
S=\int\frac{d\tau}{e}\Phi_L\left(G_{\tau m'}G_\tau\vp{G}^{m'}-\Phi_L^{-2}[(G_\tau\cdot G_y)^2-G^2_ya^2_\tau+2\Phi a_\tau(G_\tau\cdot G_y)]-E_{\tau a}E_\tau\vp{E}^a\right).
\eeq
Since the mass-shell condition following upon the action variation on the Lagrange multiplier $e(\tau)$ is irrelevant to establishing integrability of other equations of motion we set it to unity. Let us also note that one could redefine the Lagrange multiplier $e(\tau)$ to 'absorb' the overall factor of $\Phi_L$. Corresponding expression for the Lax pair component $\mathscr L$ follows by putting $\Phi_L=1$ in (\ref{lso23def})-(\ref{lfermidef}).

As the Lax pair encoding the $OSp(4|6)/(SO(1,3)\times U(3))$ superparticle equations is expressed in terms of the $osp(4|6)$ Cartan forms \cite{U-NPB} it is helpful to expand $G^{m'}(d)$ on them and $d\upsilon$
\beq\label{expan1}
\begin{array}{rl}
G^{m'}(d)&=G^{0'n}(d)M_n{}^{m'}+G^{3n}(d)N_n{}^{m'}+\Delta(d)L^{m'}+G^{kl}(d)K_{kl}{}^{m'}\\[0.2cm]
+&q^{m'\mu}d\theta_\mu+\bar q^{m'\mu}d\bar\theta_\mu+s^{m'\mu}d\eta_\mu+\bar s^{m'\mu}d\bar\eta_\mu,\\[0.2cm]
E_a(d)&=i\Omega_a(d)+u_{(1)}\vp{u}^\mu\omega_{(1)\mu a}(d)+u_{(3)}\vp{u}^\mu\omega_{(3)\mu a}(d),\\[0.2cm]
E^a(d)&=i\Omega^a(d)+\bar u_{(1)}\vp{\bar u}^\mu\bar\omega_{(1)}\vp{\bar\omega}^a_\mu(d)+\bar u_{(3)}\vp{\bar u}^\mu\bar\omega_{(3)}\vp{\bar\omega}^a_\mu(d).
\end{array}
\eeq
Coefficients at the differentials of the 'broken' fermions $d\upsilon$ are nothing but the $AdS_4\times S^7$ supervielbein components while those at the $osp(4|6)$ Cartan forms can be named 'previelbeins'. All of them are functions of 8 fermionic coordinates for the broken supersymmetries only. Analogous expansion for $a(d)$ is
\beq\label{expan2}
\begin{array}{rl}
a(d)=&\widetilde\Omega_a{}^a(d)+G^{0'm}(d)m_{m}+G^{3m}(d)n_{m}+\Delta(d)l+G^{mn}(d)k_{mn}\\[0.2cm]
+&h^\mu d\theta_\mu+\bar h^\mu d\bar\theta_\mu+p^\mu d\eta_\mu+\bar p^\mu d\bar\eta_\mu.
\end{array}
\eeq
Coefficients $h^\mu$, $p^\mu$ and c.c. can be identified with the corresponding $AdS_4\times S^7$ supervielbein components and similarly $m_{m}$, $n_{m}$, $l$ and $k_{mn}$ can be considered as 'previelbeins'.

For practical calculations explicit form of the above introduced expansion coefficients is required that can be derived upon specifying $\mathscr G_{\mathrm{br}}$ in (\ref{scoset}). We concentrate on the same representative used in our previous studies \cite{U09}, \cite{U-IJMPA}\footnote{Let us note that fermionic coordinates associated with the generators of superconformal algebras were introduced in \cite{9807115}, \cite{Kallosh2}, \cite{PST}, \cite{MT2000} to obtain explicit form of the Lagrangians of string/brane models related to the maximally supersymmetric instances of $AdS/CFT$ correspondence \cite{Maldacena}.}
\beq
\mathscr G_{\mathrm{br}}=e^{\theta^\mu Q^4_\mu+\bar\theta^\mu\bar Q_{\mu 4}}e^{\eta_\mu S^{\mu 4}+\bar\eta_\mu\bar S^\mu_4}.
\eeq
Then the 'previelbein' coefficients in (\ref{expan1}) take the form
\beq
\begin{array}{rl}
M_n{}^m=&\delta_n^m\left[1-(\theta\bar\theta)(\eta\bar\eta)+\frac14(\theta^2\bar\theta^2+\eta^2\bar\eta^2)+\frac18\theta^2\bar\theta^2\eta^2\bar\eta^2\right]-i(\theta\sigma_n\tilde\sigma^m\bar\eta+\bar\theta\sigma_n\tilde\sigma^m\eta)\\[0.2cm]
+&2\left\{1-\frac{i}{2}[(\theta\bar\eta)+(\bar\theta\eta)]\right\}(\theta\sigma_n\bar\theta)(\eta\sigma^m\bar\eta),\quad
M_n{}^3=[(\theta\bar\eta)-(\bar\theta\eta)](\theta\sigma_n\bar\theta),\\[0.2cm]
N_n{}^m=&\delta_n^m\left[-(\theta\bar\theta)(\eta\bar\eta)+\frac14(\theta^2\bar\theta^2-\eta^2\bar\eta^2)+\frac18\theta^2\bar\theta^2\eta^2\bar\eta^2\right]-i(\theta\sigma_n\tilde\sigma^m\bar\eta+\bar\theta\sigma_n\tilde\sigma^m\eta)\\[0.2cm]
+&2\left\{1-\frac{i}{2}[(\theta\bar\eta)+(\bar\theta\eta)]\right\}(\theta\sigma_n\bar\theta)(\eta\sigma^m\bar\eta),\quad
N_n{}^3=[(\theta\bar\eta)-(\bar\theta\eta)](\theta\sigma_n\bar\theta),\\[0.2cm]
L^m=&[(\bar\theta\eta)-(\theta\bar\eta)](\eta\sigma^m\bar\eta),\quad -L^3=1+i[(\theta\bar\eta)+(\bar\theta\eta)],\\[0.2cm]
K_{kl}{}^m=&-\frac{i}{2}\varepsilon_{kl}{}^m\left\{\left(1+\frac12\eta^2\bar\eta^2\right)(\theta\bar\theta)+(\eta\bar\eta)\right\}+\frac12[(\bar\theta\sigma_{kl}\eta)-(\theta\sigma_{kl}\bar\eta)](\eta\sigma^m\bar\eta),\\[0.2cm]
K_{kl}{}^3=&-\frac{i}{2}[(\bar\theta\sigma_{kl}\eta)+(\theta\sigma_{kl}\bar\eta)]
\end{array}
\eeq
and
\beq
u_{(1,3)}\vp{u}^{\mu}=\mp2\{\theta^\mu\pm i\eta^\mu[1\pm(\theta\bar\theta)]\},\quad\bar u_{(1,3)}^{\hp{(1,3)}\mu}=\mp2\{\bar\theta^\mu\mp i\bar\eta^\mu[1\mp(\theta\bar\theta)]\},
\eeq
and those from (\ref{expan2}) read
\beq
\begin{array}{rl}
m_{m}=&\left\{1-i[(\theta\bar\eta)+(\bar\theta\eta)]\right\}(\theta\sigma_m\bar\theta)+\left(1-\frac12\theta^2\bar\theta^2\right)(\eta\sigma_m\bar\eta),\\[0.2cm]
n_{m}=&\left\{1-i[(\theta\bar\eta)+(\bar\theta\eta)]\right\}(\theta\sigma_m\bar\theta)-\left(1+\frac12\theta^2\bar\theta^2\right)(\eta\sigma_m\bar\eta)\\[0.2cm]
l=&(\bar\theta\eta)-(\theta\bar\eta),\quad k_{mn}=\frac12[(\bar\theta\sigma_{mn}\eta)-(\theta\sigma_{mn}\bar\eta)]-i(\theta\bar\theta)(\eta\sigma_{mn}\bar\eta).
\end{array}
\eeq
Non-zero $AdS_4\times S^7$ supervielbein components that enter (\ref{expan1}) and (\ref{expan2}) are given by
\beq
\begin{array}{rl}
q^{m\mu}=&\frac{i}{2}\left(1+\frac12\eta^2\bar\eta^2\right)\tilde\sigma^{m\mu\nu}\bar\theta_\nu+\frac12\bar\eta^2\tilde\sigma^{m\mu\nu}\eta_\nu,\quad q^{3\mu}=-i\bar\eta^\mu,\\[0.2cm]
s^{m\mu}=&\frac{i}{2}\tilde\sigma^{m\mu\nu}\bar\eta_\nu,\quad h^\mu=-\bar\eta^\mu+i(\bar\theta\eta)\bar\eta^\mu-i(\bar\theta\bar\eta)\eta^\mu
\end{array}
\eeq and c.c. expressions. Finally proportional to $dy$
contributions to the $AdS_4\times S^7$ supervielbein components in
directions tangent to Anti-de Sitter space-time and $S^1$ fiber
acquire the form \beq
\begin{array}{rl}
G_y^m=&2\left(1+\frac12\eta^2\bar\eta^2\right)(\theta\sigma^m\bar\theta)+2\{1-i[(\theta\bar\eta)+(\bar\theta\eta)]\}(\eta\sigma^m\bar\eta),\quad G_y^3=2[(\theta\bar\eta)-(\bar\theta\eta)], \\[0.2cm]
\Phi=&1-2i[(\theta\bar\eta)+(\bar\theta\eta)]+4[(\theta\eta)(\bar\theta\bar\eta)-(\theta\bar\eta)(\bar\theta\eta)].
\end{array}
\eeq

Superparticle equations of motion can be written in the form facilitating their Lax representation. Taking $osp(4|6)/(so(1,3)\times u(3))$ Cartan forms (\ref{gradedcf}) as independent variation parameters yields the set of bosonic and fermionic equations of motion
\beq\label{beom}
\begin{array}{rl}
-\frac{\delta S}{\delta G^{0'}\vp{G}_m(\delta)}=&\dot a^{0'm}+2G_{\tau\hp{m}n}^{\hp{\tau}m}a^{0'n}-4l^{mn}G_{\tau\hp{0'}n}^{\hp{\tau}0'}+4fG_\tau\vp{G}^{3m}-2\Delta_\tau b^{3m}\\[0.2cm]
-&4i\left(\omega_{(1)\tau a}\sigma^m\bar\varepsilon_{(1)}{}^{a}-\varepsilon_{(1)a}\sigma^m\bar\omega_{(1)\tau}{}^a+\omega_{(3)\tau a}\sigma^m\bar\varepsilon_{(3)}{}^{a}-\varepsilon_{(3)a}\sigma^m\bar\omega_{(3)\tau}{}^a\right)=0,\\[0.2cm]
-\frac12\frac{\delta S}{\delta\Delta(\delta)}=&\dot f+G_{\tau\hp{0'}m}^{\hp{\tau}0'}b^{3m}-G_{\tau3m}a^{0'm}\\[0.2cm]
-&2\left(\omega_{(1)}\vp{\omega}^{\vp{\mu}}_\tau\vp{omega}^\mu_{a}\bar\varepsilon_{(1)}{}^{a}_\mu-\varepsilon_{(1)}{}^\mu_{a}\bar\omega_{(1)}\vp{\bar\omega}_\tau^{\vp{a}}\vp{\bar\omega}^a_\mu-\omega_{(3)}\vp{\omega}_\tau^{\vp{\mu}}\vp{omega}^\mu_a\bar\varepsilon_{(3)}{}^{a}_\mu+\varepsilon_{(3)}{}^\mu_{a}\bar\omega_{(3)}\vp{\bar\omega}_\tau^{\vp{a}}\vp{\bar\omega}^a_\mu\right)=0,\\[0.2cm]
-\frac{\delta S}{\delta\Omega_a(\delta)}=&\dot y^a+iy^b\left(\widetilde\Omega_{\tau b}\vp{\Omega}^a+\delta_b^a\widetilde\Omega_{\tau c}\vp{\Omega}^c\right)-4iw\Omega_\tau\vp{\Omega}^a\\[0.2cm]
+&4i\varepsilon^{abc}\left(\omega_{(1)}\vp{\omega}_\tau^{\vp{\mu}}\vp{omega}^\mu_b\varepsilon_{(1)\mu c}-\omega_{(3)}\vp{\omega}_\tau^{\vp{\mu}}\vp{omega}^\mu_b\varepsilon_{(3)\mu c}\right)=0
\end{array}
\eeq
and
\begin{equation}\label{feom}
\begin{array}{rl}
\frac14\frac{\delta S}{\delta\bar\omega_{(1)}\vp{\bar\omega}^a_\mu(\delta)}=&\dot\varepsilon_{(3)}{}^\mu_{a}\!+\!\frac12\!\left(G_\tau{}^{mn}\varepsilon_{(3)}{}^\nu_{a}\!-\! l^{mn}\omega_{(3)}\vp{\omega}_{\tau a}^{\hp{\tau}\nu}\right)\!\sigma_{mn\nu}{}^\mu\!+\! i\tilde\sigma_m^{\mu\nu}\!\left(G_\tau\vp{G}^{3m}\varepsilon_{(3)\nu a}\!-\!\frac{1}{2}b^{3m}\omega_{(3)}\vp{\omega}_{\tau\nu a}\right)\!\\[0.2cm]
+&i\tilde\sigma^{\mu\nu}_m\left(G_\tau\vp{G}^{0'm}\varepsilon_{(1)\nu a}-\frac{1}{2}a^{0'm}\omega_{(1)}\vp{\omega}_{\tau\nu a}\right)+\Delta_\tau\varepsilon_{(1)}{}^\mu_{a}- f\omega_{(1)}\vp{\omega}^{\hp{\tau}\mu}_{\tau a}\\[0.2cm]
-&i\!\left(\widetilde\Omega_{\tau a}\vp{\Omega}^b\!-\!\delta^b_a\widetilde\Omega_{\tau c}\vp{\Omega}^c\right)\!\varepsilon_{(3)}{}^\mu_{b}\!-\!2iw\omega_{(3)}\vp{\omega}^{\hp{\tau}\mu}_{\tau a}\!-\! i\varepsilon_{abc}\left(\Omega_\tau{}^b\bar\varepsilon_{(1)}{}^{\mu c}\!-\! y^b\bar\omega_{(1)}\vp{\bar\omega}_\tau^{\hp{\tau}\mu c}\right)=0,\\[0.2cm]
-\frac14\frac{\delta S}{\delta\bar\omega_{(3)}\vp{\bar\omega}^a_\mu(\delta)}=&\dot\varepsilon_{(1)}{}^\mu_{a}\!+\!\frac12\!\left(G_\tau{}^{mn}\varepsilon_{(1)}{}^\nu_{a}\!-\! l^{mn}\omega_{(1)}\vp{\omega}_{\tau a}^{\hp{\tau}\nu}\right)\!\sigma_{mn\nu}{}^\mu\!-\! i\tilde\sigma_m^{\mu\nu}\!\left(G_\tau{}^{3m}\varepsilon_{(1)\nu a}\!-\!\frac{1}{2}b^{3m}\omega_{(1)}\vp{\omega}_{\tau\nu a}\right)\!\\[0.2cm]
-&i\tilde\sigma^{\mu\nu}_m\left(G_\tau\vp{G}^{0'm}\varepsilon_{(3)\nu a}-\frac{1}{2}a^{0'm}\omega_{(3)}\vp{\omega}_{\tau\nu a}\right)+\Delta_\tau\varepsilon_{(3)}{}^\mu_{a}-f\omega_{(3)}\vp{\omega}^{\hp{\tau}\mu}_{\tau a}\\[0.2cm]
-&i\!\left(\widetilde\Omega_{\tau a}\vp{\Omega}^b-\delta^b_a\widetilde\Omega_{\tau c}\vp{\Omega}^c\right)\!\varepsilon_{(1)}{}^\mu_{b}\!-\!2iw\omega_{(1)}\vp{\omega}^{\hp{\tau}\mu}_{\tau a}\!-\! i\varepsilon_{abc}\left(\Omega_\tau{}^b\bar\varepsilon_{(3)}{}^{\mu c}\!-\! y^b\bar\omega_{(3)}\vp{\bar\omega}_{\tau}^{\hp{\tau}\mu c}\right)=0
\end{array}
\end{equation}
and c.c. equations. In (\ref{beom}) and (\ref{feom}) we have introduced the following bosonic and fermionic quantities
\beq\label{lso23def} 
\!\left(\!\!
\begin{array}{c}
\! a^{0'}_{\hp{0'}m}\!\!\\
\!2f\!\!\\
\!-b_{3m}\!\!\\
\!-l_{kl}\!\!
\end{array}
\!\!\right)\!=2\!\left(\!\!
\begin{array}{c}
\! M_m{}^{n'}\!\\
\! L^{n'}\!\\
\! N_m{}^{n'}\!\\
\! K_{kl}\vp{K}^{n'}\! 
\end{array}
\!\!\right)\![\Phi_LG_{\tau n'}\!-\!\Phi^{-1}_L(G_\tau\!\cdot\! G_y\!+\!\Phi a_\tau)G_{yn'}]\!+\!2\Phi^{-1}_L\!\left(\!
\begin{array}{c}
\! m_{m}\!\!\\
\! l\!\!\\
\! n_{m}\!\!\\
\! k_{kl}\!\!
\end{array}
\!\right)\!(G^2_ya_\tau\!-\!\Phi G_\tau\!\cdot\! G_y), 
\eeq 
\beq\label{lsu4def}  
y^a=-i\Phi_LE_\tau{}^a,\quad\bar y_a=-i\Phi_LE_{\tau a},\quad w=\frac12\Phi^{-1}_L(G^2_ya_\tau-\Phi G_\tau\cdot G_y).
\eeq 
and
\beq\label{lfermidef}
\begin{array}{c}
\varepsilon_{(1)}{}^\mu_{a}=-\frac14\Phi_LE_{\tau a}\bar u_{(3)}\vp{\bar u}^\mu,\quad\varepsilon_{(3)}{}^\mu_{a}=\frac14\Phi_LE_{\tau a}\bar u_{(1)}\vp{\bar u}^\mu,\\[0.2cm]
\bar\varepsilon_{(1)}{}^{\mu a}=\frac14\Phi_LE_\tau{}^au_{(3)}\vp{u}^\mu,\quad
\bar\varepsilon_{(3)}{}^{\mu a}=-\frac{1}{4}\Phi_LE^{\vp{a}}_{\tau}\vp{E}^a_{\vp{tau}}u_{(1)}\vp{u}^\mu
\end{array}
\eeq
that enter the Lax component $\mathscr L$ (\ref{lax2}). In addition there are 8 equations stemming from the variation of the action (\ref{action}) on $\upsilon$. They can be brought to the following form
\beq\label{brfeom}
\begin{array}{c}
-2\frac{d}{d\tau}\left[\Phi_LG_{\tau m'}\frac{\partial G_\tau\vp{G}^{m'}}{\partial\dot\upsilon_\mu}-\Phi_L^{-1}(G_\tau\!\cdot\! G_y\!+\! \Phi a_\tau)G_{ym'}\frac{\partial G_\tau\vp{G}^{m'}}{\partial\dot\upsilon_\mu}+\Phi_L^{-1}(G^2_ya_\tau\!-\!\Phi G_\tau\!\cdot\! G_y)\frac{\partial a_\tau}{\partial\dot\upsilon_\mu}\right]\\[0.2cm]
+\frac{\partial L}{\partial\Phi_L}\frac{\partial\Phi_L}{\partial\upsilon_\mu}+\Phi_L^{-1}a_\tau\left(a_\tau\frac{\partial G^2_y}{\partial\upsilon_\mu}-2G_\tau\!\cdot\! G_y\frac{\partial\Phi}{\partial\upsilon_\mu}\right)+2\Phi_LG_{\tau m'}\left(G_\tau\vp{G}^{0'n}\frac{\partial M_n{}^{m'}}{\partial\upsilon_\mu}+G_\tau\vp{G}^{3n}\frac{\partial N_n{}^{m'}}{\partial\upsilon_\mu}\right.\\[0.2cm]
\left.+\Delta_\tau\frac{\partial L^{m'}}{\partial\upsilon_\mu}+G_\tau\vp{G}^{kl}\frac{\partial K_{kl}\vp{K}^{m'}}{\partial\upsilon_\mu}
-\dot\theta_\nu\frac{\partial q^{m'\nu}}{\partial\upsilon_\mu}-\dot{\bar\theta}_\nu\frac{\partial\bar q^{m'\nu}}{\partial\upsilon_\mu}-\dot\eta_\nu\frac{\partial s^{m'\nu}}{\partial\upsilon_\mu}-\dot{\bar\eta}_\nu\frac{\partial\bar s^{m'\nu}}{\partial\upsilon_\mu}\right)\\[0.2cm]
+\Phi_L\left[E_{\tau a}\left(\bar\omega_{(1)}\vp{\bar\omega}_{\tau\nu}^{\hp{\tau}a}\frac{\partial\bar u_{(1)}\vp{\bar u}^\nu}{\partial\upsilon_\mu}+\bar\omega_{(3)}\vp{\bar\omega}_{\tau\nu}^{\hp{\tau}a}\frac{\partial\bar u_{(3)}\vp{\bar u}^\nu}{\partial\upsilon_\mu}\right)+E_\tau^{\hp{\tau}a}\left(\omega_{(1)}\vp{\omega}_{\tau\nu a}\frac{\partial u_{(1)}\vp{u}^\nu}{\partial\upsilon_\mu}+\omega_{(3)}\vp{\omega}_{\tau\nu a}\frac{\partial u_{(3)}\vp{u}^\nu}{\partial\upsilon_\mu}\right)\right]\\[0.2cm]
\!-2\Phi_L^{-1}(G_\tau\!\cdot\! G_y\!+\!\Phi a_\tau)\!\left(\! G_\tau^{\hp{\tau}0'm}\frac{\partial M_m{}^{n'}G_{yn'}}{\partial\upsilon_\mu}\!+\! G_\tau^{\hp{\tau}3m}\frac{\partial N_m{}^{n'}G_{yn'}}{\partial\upsilon_\mu}\!+\!\Delta_\tau\frac{\partial L^{m'}G_{ym'}}{\partial\upsilon_\mu}\!+\! G_\tau^{\hp{\tau}kl}\frac{\partial K_{kl}^{\hp{kl}n'}G_{yn'}}{\partial\upsilon_\mu}\right.\\[0.2cm]
-\left.\dot\theta_\nu\frac{\partial q^{m'\nu}G_{ym'}}{\partial\upsilon_\mu}-\dot{\bar\theta}_\nu\frac{\partial\bar q^{m'\nu}G_{ym'}}{\partial\upsilon_\mu}-\dot\eta_\nu\frac{\partial s^{m'\nu}G_{ym'}}{\partial\upsilon_\mu}-\dot{\bar\eta}_\nu\frac{\partial\bar s^{m'\nu}G_{ym'}}{\partial\upsilon_\mu}\right)\\[0.2cm]
+2\Phi_L^{-1}(G^2_ya_\tau-\Phi G_\tau\!\cdot\! G_y)(G_\tau^{\hp{\tau}0'm}\frac{\partial m_{m}}{\partial\upsilon_\mu}+G_\tau^{\hp{\tau}3m}\frac{\partial n_{m}}{\partial\upsilon_\mu}+\Delta_\tau\frac{\partial l}{\partial\upsilon_\mu}+G_\tau^{\hp{\tau}mn}\frac{\partial k_{mn}}{\partial\upsilon_\mu}\\[0.2cm]
-\dot\theta_\nu\frac{\partial h^\nu}{\partial\upsilon_\mu}-\dot{\bar\theta}_\nu\frac{\partial\bar h^\nu}{\partial\upsilon_\mu}-\dot\eta_\nu\frac{\partial p^\nu}{\partial\upsilon_\mu}-\dot{\bar\eta}_\nu\frac{\partial\bar p^\nu}{\partial\upsilon_\mu})=0,
\end{array}
\eeq
where for the sake of uniformity of presentation we introduced the following notation $\frac{\partial G_\tau\vp{G}^{m'}}{\partial\dot\upsilon_\mu}=(q^{m'\mu}, \bar q^{m'\mu}, s^{m'\mu}, \bar s^{m'\mu})$ and $\frac{\partial a_\tau}{\partial\dot\upsilon_\mu}=(h^\mu, \bar h^\mu, p^\mu, \bar p^\mu)$.\footnote{We assume that the fermionic derivative acts from the right.}

Fulfilment of the $AdS_4\times\mathbb{CP}^3$ superparticle equations of motion (\ref{beom}), (\ref{feom}), (\ref{brfeom}) is equivalent to the Lax equation (\ref{lax1}) with the Lax component $\mathscr M$ given by the world-line projection of the $osp(4|6)$ Cartan forms
(\ref{gradedcf}) and another component can be presented as
\beq\label{lax2}
\mathscr L=\mathscr L_{\mbox{\scriptsize so(2,3)}}+\mathscr L_{\mbox{\scriptsize su(4)}}+\mathscr L_{\mbox{\scriptsize24susys}}\in osp(4|6).
\eeq
The first term takes value in the $so(2,3)$ isometry algebra of $AdS_4$ space-time
\beq
\mathscr L_{\mbox{\scriptsize so(2,3)}}=a^{0'm}M_{0'm}+fD+b^{3m}M_{3m}+l^{mn}M_{mn}.
\eeq
The second summand in (\ref{lax2}) belongs to the $su(4)$ isometry algebra of $\mathbb{CP}^3$ manifold
\beq
\mathscr L_{\mbox{\scriptsize su(4)}}=y^aT_a+\bar y_{a}T^a+4wV_a{}^a
\eeq
and the last one is the linear combination of the fermionic generators
\beq
\mathscr L_{\mbox{\scriptsize24susys}}=\varepsilon_{(1)}{}^\mu_{a}Q^{\vp{a}}_{(1)}\vp{Q}^a_\mu+\bar\varepsilon_{(1)}{}^{\mu a}\bar Q_{(1)\mu a}+\varepsilon_{(3)}{}^\mu_{a}Q^{\vp{a}}_{(3)}\vp{Q}^a_\mu+\bar\varepsilon_{(3)}{}^{\mu a}\bar Q_{(3)\mu a}.
\eeq
The Lax component $\mathscr L$ can be presented in the form of $osp(4|6)-$valued differential operator (\ref{lax2compl}) acting of the superparticle action (\ref{action}).

\section{Conclusion}
In this paper generalizing the consideration of Ref.~\cite{U-NPB} we have shown that the complete set of equations of motion for the massless $AdS_4\times\mathbb{CP}^3$ superparticle admits Lax representation implying their classical integrability. It would be of interest to construct explicitly all the integrals of motion as well as to examine quantum integrability of this model. It is also tempting to suggest that the results reported here may be useful in proving integrability of the $AdS_4\times\mathbb{CP}^3$ superstring and $D0-$brane equations that can be reduced to those of the massless superparticle.

\section{Acknowledgements}

The author is grateful to A.A.~Zheltukhin for stimulating discussions.

\end{document}